\documentclass[conference]{IEEEtran}
\usepackage{url}
\usepackage{xspace}
\usepackage{subfigure}
\usepackage{graphicx}
\usepackage{algorithm}
\usepackage{algorithmicx,xspace,algpseudocode}
\usepackage{listings}
\usepackage{pbox}
\usepackage{color}
\usepackage{cuted}
\makeatletter
\g@addto@macro{\UrlBreaks}{\UrlOrds}
\makeatother
\usepackage{amsmath}
\usepackage{amssymb}
\usepackage[noadjust]{cite}

\algrenewcommand{\algorithmiccomment}[1]{$\vartriangleright$ #1}

\def\keywords{\vspace{.5em}
{\textit{Keywords}:\,\relax%
}}

\ifCLASSINFOpdf
\else
\fi

\begin{document}
\title{Nonlinear Unknown Input and State Estimation Algorithm in Mobile Robots\\
\large Technical Report No. Cyber-Security-Lab-2018-001}

\author{\IEEEauthorblockN{Pinyao Guo\IEEEauthorrefmark{1},
Hunmin Kim\IEEEauthorrefmark{2},
Nurali Virani\IEEEauthorrefmark{3}, 
Jun Xu\IEEEauthorrefmark{1}, 
Minghui Zhu\IEEEauthorrefmark{2} and
Peng Liu\IEEEauthorrefmark{1}}
\IEEEauthorblockA{\IEEEauthorrefmark{1}College of Information Sciences and Technology, Pennsylvania State University, University Park, PA 16802, USA\\
\texttt{\{pug132,jxx13,pliu\}@ist.psu.edu}}
\IEEEauthorblockA{\IEEEauthorrefmark{2}School of Electrical Engineering and Computer Science, Pennsylvania State University, University Park, PA 16802, USA\\
\texttt{\{huk164,muz16\}@psu.edu}}
\IEEEauthorblockA{\IEEEauthorrefmark{3}GE Global Research, Niskayuna, NY 12309, USA\\
\texttt{nurali.virani@ge.com}}}


\maketitle

\begin{abstract}
This technical report provides the description and the derivation of a novel nonlinear unknown input and state estimation algorithm (NUISE) for mobile robots. The algorithm is designed for real-world robots with nonlinear dynamic models and subject to stochastic noises on sensing and actuation. Leveraging sensor readings and planned control commands, the algorithm detects and quantifies anomalies on both sensors and actuators. Later, we elaborate the dynamic models of two distinctive mobile robots for the purpose of demonstrating the application of NUISE. This report serves as a supplementary document for \cite{dsn_paper}.

\end{abstract}

% \fi
\IEEEpeerreviewmaketitle

\section{NUISE Algorithm and its Derivation}\label{sec:AP_n}
%NUISE is the essential part of the robot intrusion detection algorithm~\eqref{nalgo1} where it is recursively called.
%The NISE takes inputs: on-board sensor outputs $\textbf{z}_k$, control input $\textbf{u}_k$, previous state estimate $\hat{\textbf{x}}_{k-1|k-1}$, previous error covariance $P_{k-1}^x$.
%We present the derivation of the NUISE algorithm.
%The NUISE produces state and actuator attack estimates, and their error covariance matices where the error covariance represents how the estimated state is accurate. 
%Two important criteria are applied to the derivation of the NUISE. That is, the estimate is unbiased and its error covariance is minimized. We use a linear estimator gain called the best linear unbiased estimator (BLUE)
Minimum variance unbiased state and unknown input estimation is first introduced in~\cite{kitanidis1987unbiased} with indirect feedthrough\footnote{Indirect feedthrough suggests that the input of a system indirectly influences the output through system states change. Direct feedthrough suggests that the input of a system is directly connected/fed to the output.} unknown input. The method has been extended by many research studies.
A general parameterized gain matrix is derived in~\cite{darouach1997unbiased}. Estimation with direct feedthrough unknown input is proposed in~\cite{cheng2009unbiased,hou1998optimal}.
Young et al.~\cite{SY-MZ-EF:Automatica16} analyze
the stability of systems with direct and indirect feedthrough unknown input.
%by generalizing the solution~\cite{darouach1997unbiased}, integrating directly measurable unknown input~\cite{cheng2009unbiased,hou1998optimal}, and analyzing the stability~\cite{SY-MZ-EF:Automatica16}.
Estimators with indirect feedthrough unknown input has been applied to the fault detection in systems without noise~\cite{chen1996design} and with noises~\cite{de1997nonparametric,liu2011robust}.
An estimator with both direct and indirect feedthrough unknown input is proposed~\cite{yong2016simultaneous} for the attack detection in systems with noises, where the attack location is unknown.

One limitation of the aforementioned works is that the proposed methods are limited to handle linear systems. An estimator that can handle nonlinear systems is unexplored. In this work, we propose the nonlinear unknown input and state estimation algorithm (NUISE) as an extension of the above references for nonlinear systems. The algorithm can also be viewed as an extension of 
the extended Kalman filters~\cite{jazwinski2007stochastic} for state estimation of nonlinear systems by integrating unknown input estimation.
It is the first time to study the state and unknown input estimation problem in stochastic nonlinear systems. Leveraging the reference sensor readings and planned control commands from the last iteration, NUISE estimates new robot states, corruptions in testing sensor readings, corruptions in control commands, and a likelihood for each mode.

Algorithm~\ref{nalgo2_full} describes the complete NUISE algorithm. We first present the definition of optimal estimates in an estimation problem.
Optimal estimates contain two properties. Firstly, the estimates are unbiased, i.e., its expected value is equal to the targeted value. Secondly, the estimates have a minimum error covariance matrix, i.e., the estimation error variances are minimized with the given information.
%We call such optimal estimate as the best linear unbiased estimator (BLUE).

\begin{algorithm}
\small
\caption{Nonlinear Unknown Input and State Estimation Algorithm (NUISE)} \label{nalgo2_full}
\begin{algorithmic}[1]
\Require $\textbf{u}_{k-1}$, $\hat{\textbf{x}}_{k-1|k-1}$, $\textbf{z}_{1,k}$, $\textbf{z}_{2,k}$%, $P_{k-1}^x$
\Ensure $\hat{\textbf{x}}_{k|k}$, $\hat{\textbf{d}}_{k}^{s}$, $\hat{\textbf{d}}_{k-1}^{a}$, ${\mathcal N}_k$%$P_{k}^{x}$, $P_{k}^{s}$, $P_{k-1}^{a}$, 
\State Initialize;\par
\hspace{-5mm}\Comment{\textbf{Actuator anomaly vector $\textbf{d}_{k-1}^{a}$ estimation}}
\State $\tilde{P}_{k-1} \leftarrow A_{k-1} P_{k-1}^x(A_{k-1})^T+Q_{k-1}$;
\State $\tilde{R}_{2,k}^{*} \leftarrow C_{2,k} \tilde{P}_{k-1}(C_{2,k})^T+R_{2,k}$;
\State $M_{2,k}\leftarrow((G_{k-1})^T(C_{2,k})^T(\tilde{R}_{2,k}^{*})^{-1}C_{2,k}G_{k-1})^{-1}$ $(G_{k-1})^T(C_{2,k})^T(\tilde{R}_{2,k}^{*})^{-1}$;
\State $\hat{\textbf{d}}_{k-1}^{a} \leftarrow M_{2,k}(\textbf{z}_{2,k}-C_{2,k}f(\hat{\textbf{x}}_{k-1|k-1},\textbf{u}_{k-1}))$;
\State $P_{k-1}^{a}\leftarrow M_{2,k}\tilde{R}_{2,k}^{*}(M_{2,k})^T$;\par
\hspace{-5mm}\Comment{\textbf{State prediction}}
\State $\hat{\textbf{x}}_{k|k-1} \leftarrow f(\hat{\textbf{x}}_{k-1|k-1},\textbf{u}_{k-1}+\hat{\textbf{d}}_{k-1}^{a})$;
\State $\bar{A}_{k-1}\leftarrow (I-G_{k-1}M_{2,k}C_{2,k}) A_{k-1}$;
\State %$\bar{Q}_{k-1} \triangleq {\mathbb E}[\bar{\zeta}_{k-1}\bar{\zeta}_{k-1}^T]$ with $\bar{\zeta}_{k-1} \triangleq -(I-G_{2,k-1}M_{2,k}C_{2,k})(G_{1,k-1}M_{1,k-1}\xi_{1,k-1}+\zeta_{k-1})-G_{2,k-1}M_{2,k}\xi_{2,k}$;
$\bar{Q}_{k-1} \leftarrow (I-G_{k-1}M_{2,k}C_{2,k})Q_{k-1}(I-G_{k-1}M_{2,k}C_{2,k})^T+G_{k-1}M_{2,k}R_{2,k}(M_{2,k})^T(G_{k-1})^T$;
\State $P_{k|k-1}^{x}\leftarrow\bar{A}_{k-1}P_{k-1}^{x}(\bar{A}_{k-1})^T+\bar{Q}_{k-1}$;\par
%$+\bar{A}_{k-1}R_{2,k}M_{2,k}^TG_{2,k-1}^T+G_{2,k-1}M_{2,k}R_{2,k}\bar{A}_{k-1}^T$;
\hspace{-5mm}\Comment{\textbf{State estimation}}
\State $\tilde{R}_{2,k} \leftarrow C_{2,k}P_{k|k-1}^{x}(C_{2,k})^T+R_{2,k} + C_{2,k}G_{k-1}M_{2,k}R_{2,k}+R_{2,k}(M_{2,k})^T(G_{k-1})^T(C_{2,k})^T$;
\State $L_k\leftarrow(C_{2,k}P_{k|k-1}^{x}+R_{2,k}(M_{2,k})^T(G_{k-1})^T)^T(\tilde{R}_{2,k})^{-1}$;
%\State $L_k=P_{k|k-1}^xC_k^T \tilde{R}_{2,k}^{-1}$ where $\tilde{R}_{2,k} \triangleq C_{2,k}P_{k|k-1}^xC_{2,k}^T+R_{2,k}$;
\State $\hat{\textbf{x}}_{k|k} \leftarrow \hat{\textbf{x}}_{k|k-1}+L_k (\textbf{z}_{2,k}-h_{2}(\hat{\textbf{x}}_{k|k-1}))$;
\State $P_{k}^{x}\leftarrow(I-L_kC_{2,k})P_{k|k-1}^{x}(I-L_kC_{2,k})^T + L_k R_{2,k} (L_k)^T-(I-L_k C_{2,k}) G_{k-1}M_{2,k}R_{2,k}(L_k)^T-L_kR_{2,k}(M_{2,k})^T(G_{k-1})^T(I-L_k C_{2,k})^T$;\par
\hspace{-5mm}\Comment{\textbf{Sensor anomaly vector $\textbf{d}_{k}^{s}$ estimation}}
\State $\hat{\textbf{d}}_{k}^{s}\leftarrow \textbf{z}_{1,k}-h_{1}(\hat{\textbf{x}}_{k|k})$; 
\State $P_{k}^{s}\leftarrow C_{1,k}P_{k}^{x} (C_{1,k})^T + R_{1,k}$;\par
\hspace{-5mm}\Comment{\textbf{Likelihood of the mode}}
\State $\nu_k \leftarrow \textbf{z}_{2,k} -h_{2}(\hat{\textbf{x}}_{k|k-1})$;
\State $\bar{P}_{k|k-1} \leftarrow C_{2,k}P_{k|k-1}^{x}(C_{2,k})^T + R_{2,k}-C_{2,k}G_{k-1}M_{2,k}R_{2,k}-R_{2,k}(M_{2,k})^T(G_{k-1})^T(C_{2,k})^T$;
\State $n\leftarrow rank(\bar{P}_{k|k-1})$;
\State ${\mathcal N}_k \leftarrow \frac{1}{(2 \pi)^{n/2} |\bar{P}_{k|k-1}|_{+}^{1/2}}\exp(-\frac{(\nu_k)^T(\bar{P}_{k|k-1})^{\dagger}\nu_k}{2})$;\protect\footnotemark
\end{algorithmic}
\end{algorithm}
\footnotetext{Notations $\dagger$ and $|\cdot|_{+}$ refer pseudoinverse and pseudodeterminant, respectively. $n$ refers to the rank of $\bar{P}_{k|k-1}$.}

%To find an optimal estimate, we first define what is the {\color{red}meaning of optimal}. Note that, to be an accurate estimate, its expected value must be the targeted value (unbiased), and its error variance must be smaller than other estimate (minimum variance). Such intuition leads us to the best linear unbiased estimator (BLUE); i.e., unbiased estimate with minimized error covariance matrix. Linear estimation refers that the estimate is obtained by a weighted average of each sensors. Therefore, {\color{red}to find the optimal estimate at each step,} it is essential to keep track of the estimate as well as its error covariance matrice for the BLUE.

We derive the NUISE algorithm in 4 steps: 1) actuator anomaly vector estimation, 2) state prediction, 3) state estimation, and 4) testing sensor anomaly vector estimation. In each intermediate step, the estimation errors and covariance matrices are calculated accordingly in order to find the optimal estimates.

Consider a particular mode $m$ of the dynamic model (2) in~\cite{dsn_paper} with potential robot misbehaviors
\begin{align}
\textbf{x}_{k+1} &= f_k^m(\textbf{x}_k,\textbf{u}_k+\textbf{d}_{k}^{a,m})+\zeta_k^m\nonumber\\
\textbf{z}_{1,k}^m&=h_{1,k}^m(\textbf{x}_k)+\textbf{d}_{k}^{s,m}+\xi_{1,k}^m\nonumber\\
\textbf{z}_{2,k}^m&= h_{2,k}^m(\textbf{x}_k)+\xi_{2,k}^m
\label{CD001.1}
\end{align}
where vector $\textbf{d}_{k}^{s,m}$ and $\textbf{d}_{k}^{a,m}$ represent sensor anomaly vector and actuator anomaly vector, respectively.
In mode $m$, testing sensor readings $\textbf{z}_{1,k}^m$ might be modified by anomaly vector $\textbf{d}_{k}^{s,m}$. Reference sensor readings $\textbf{z}_{2,k}^m$ are assumed to be clean.
We omit mode index $m$ in the remaining part of the NUISE derivation for the ease of presentations.
The dynamic system~\eqref{CD001.1} can be linearized into
\begin{align}
\textbf{x}_{k+1} & \simeq  A_k \textbf{x}_k+ B_k \textbf{u}_k+G_{k}\textbf{d}_{k}^a +\zeta_k\nonumber\\
\textbf{z}_{1,k} & \simeq  C_{1,k} \textbf{x}_k + \textbf{d}_{k}^s+\xi_{1,k}\nonumber\\
\textbf{z}_{2,k} & \simeq  C_{2,k} \textbf{x}_k + \xi_{2,k}
\label{CD001.7}
\end{align} 
where 
\begin{align*}
&A_k \triangleq \frac{\partial f_k}{\partial \textbf{x}}\big|_{\hat{\textbf{x}}_{k|k},\textbf{u}_{k}+\hat{d}_{k-1}^a},
B_k \triangleq \frac{\partial f_k}{\partial u}\big|_{\hat{\textbf{x}}_{k|k},\textbf{u}_{k}+\hat{d}_{k-1}^a},\nonumber\\
&C_{1,k} \triangleq \frac{\partial h_{1,k}}{\partial \textbf{x}}\big|_{\hat{\textbf{x}}_{k|k-1}},
C_{2,k} \triangleq \frac{\partial h_{2,k}}{\partial \textbf{x}}\big|_{\hat{\textbf{x}}_{k|k-1}}\nonumber\\
&G_{k} \triangleq \frac{\partial f_k}{\partial \textbf{d}^a}\big|_{\hat{\textbf{x}}_{k|k},\textbf{u}_{k}+\hat{d}_{k-1}^a}.
\end{align*}

\textbf{Actuator anomaly vector $\textbf{d}_{k-1}^{a}$ estimation}: 
Given unbiased estimates of previous states $\hat{\textbf{x}}_{k-1|k-1}$, we can predict the current states using the known kinematic function $f_k(\cdot)$ as follows
\begin{align*}
\hat{\textbf{x}}_{k|k-1}^{*} = f_{k-1}(\hat{\textbf{x}}_{k-1|k-1},\textbf{u}_{k-1}).
\end{align*}
The estimation error is described as
\begin{align*}
\tilde{\textbf{x}}_{k|k-1}^* &= \textbf{x}_k-\hat{\textbf{x}}_{k|k-1}^*= A_{k-1} \tilde{\textbf{x}}_{k-1|k-1} +G_{k-1} \textbf{d}_{k-1}^a+ \nonumber\\&\zeta_{k-1}.
\end{align*}
Noticeably, the estimation is biased, i.e., ${\mathbb E}[\hat{\textbf{x}}_{k|k-1}^*] \neq \textbf{x}_{k|k-1}$, because we did not consider the possible unknown misbehaviors yet, i.e., $G_{k-1} \textbf{d}_{k-1}^a \neq 0$.
To obtain an unbiased state prediction, we needed to find the estimates of the actuator anomaly. The expected output without considering actuator misbehaviors is $C_{2,k}\hat{\textbf{x}}_{k|k-1}^*$. The discrepancy between what we expected and what we actually obtain $\textbf{z}_{2,k}-C_{2,k}\hat{\textbf{x}}_{k|k-1}^*$ indicates the impact of actuator anomaly $\textbf{d}_{k-1}^a$.
Therefore, the actuator anomaly vector estimates can be obtained linearly from the sensor output bias
\begin{align*}
\hat{\textbf{d}}_{k-1}^a &= M_{2,k}(\textbf{z}_{2,k}-C_{2,k}f_{k-1}(\hat{\textbf{x}}_{k-1|k-1},\textbf{u}_{k-1}))\nonumber\\
&=M_{2,k}(C_{2,k} A_{k-1} \tilde{\textbf{x}}_{k-1|k-1}+C_{2,k} G_{k-1}\textbf{d}_{k-1}^a\nonumber\\&+C_{2,k}\zeta_{k-1}+\xi_{2,k})
\end{align*}
where the estimator gain $M_{2,k}$ represents a weighted average of the sensor bias.
The unknown input estimates are unbiased, i.e.,
${\mathbb E}[\hat{\textbf{d}}_{k-1}^a]=\textbf{d}_{k-1}^a$ providing that ${\mathbb E}[\tilde{\textbf{x}}_{k-1|k-1}]=0$, and $M_{2,k}C_{2,k}G_{k-1}=I$.
In order to achieve optimal estimates, matrix gain $M_k$ should be carefully chosen with minimum variances. To do this, consider the sensor output bias
\begin{align*}
\tilde{\textbf{z}}_{2,k} &= C_{2,k}(G_{k-1} \textbf{d}_{k-1}^a + A_{k-1}\tilde{\textbf{x}}_{k-1|k-1}+\zeta_k)+\xi_k
\end{align*}
where ${\mathbb E}[C_{2,k}A_{k-1}\tilde{\textbf{x}}_{k-1|k-1}+C_{2,k}\zeta_{k-1}+\xi_k]=0$ and its covariances are calculated by
\begin{align*}
\tilde{R}_{2,k}^{*} &\triangleq {\mathbb E}[\tilde{\textbf{z}}_{2,k}\tilde{\textbf{z}}_{2,k}^T]=C_{2,k} \tilde{P}_{k-1}C_{2,k}^T+R_{2,k}
\end{align*}
where $\tilde{P}_k \triangleq A_{k-1} P_{k-1}^xA_{k-1}^T+Q_{k-1}$.
%which is positive definite since $P_{k-1}^x$ and $R_k$ are positive definite. Also, $Q_k$ is positive semi-definite. 
We choose the matrix $M_k$ using the Gauss Markov theorem~\cite{kailath2000linear}
\begin{align*}
M_{2,k}=(G_{k-1}^TC_{2,k}^T\tilde{R}_{2,k}^{*-1}C_{2,k}G_{k-1})^{-1}G_{k-1}^TC_{2,k}^T\tilde{R}_{2,k}^{*-1}
\end{align*}
%Since the Gauss Markov theorem requires normalized noise covariance, we scale the sensor output bias $\tilde{\textbf{z}}_k$ by $S_k^{-1}$ as follows \begin{align*} S_k^{-1}\tilde{\textbf{z}}_k &= S_k^{-1}C_kG_{k-1} \textbf{d}_{k-1} \nonumber\\ &+ S_k^{-1}(C_kA_{k-1}\tilde{\textbf{x}}_{k-1|k-1}+C_k\zeta_k+\xi_k) \end{align*} {\color{red}where the invertible matrix $S_k$ is found from $\tilde{R}_k=S_kS_k^T$.} Using Gauss Markov theorem~\cite{kailath2000linear}, a BLUE can be found by
which satisfies $M_{2,k}C_{2,k}G_{k-1}=I$. We assume that $G_{k-1}^TC_{2,k}^T\tilde{R}_{2,k}^{*-1}C_{2,k}G_{k-1}$ is invertible. Anomaly vector estimation error covariances are
$P_{k-1}^{a} \triangleq {\mathbb E}[\tilde{\textbf{d}}_{k-1}^a(\tilde{\textbf{d}}_{k-1}^a)^T]=M_{2,k}\tilde{R}_{2,k}^*M_{2,k}^T$.

\textbf{State prediction}: Estimates $\hat{\textbf{x}}_{k|k-1}^*$ are calculated under a partial knowledge of misbehaviors. Since we have the actuator anomaly estimates $\hat{\textbf{d}}_{k-1}^a$ from the previous step, we can update the state estimates
\begin{align*}
\hat{\textbf{x}}_{k|k-1}&=f_{k-1}(\hat{\textbf{x}}_{k-1|k-1},\textbf{u}_{k-1}+\hat{\textbf{d}}_{k-1}^a)
\end{align*}
The state estimates are now unbiased, i.e., ${\mathbb E}[\hat{\textbf{x}}_{k|k}]=\textbf{x}_k$, since ${\mathbb E}[\hat{\textbf{d}}_{k-1}^a]=\textbf{d}_{k-1}^a$.
Now we find the state prediction error covariance matrix
\begin{align}
P_{k|k-1}^x&=\bar{A}_{k-1}P_{k-1}^x\bar{A}_{k-1}^T+\bar{Q}_{k-1}
\label{P_kk_1}
\end{align}
where $\bar{A}_{k-1}=(I-G_{k-1}M_{2,k}C_{2,k})A_{k-1}$
and $\bar{Q}_{k-1}^j = (I-G_{k-1}M_{2,k}C_{2,k})Q_{k-1}(I-G_{k-1}M_{2,k}C_{2,k})^T+G_{k-1}M_{2,k}R_{2,k}M_{2,k}^TG_{k-1}^T$.

\textbf{State estimation}: 
Predicted states $\hat{\textbf{x}}_{k|k-1}$ are not perfect because of process and measurement noises. In order to obtain the estimates accurately considering noises, we do corrections on the state estimates using sensor readings. We utilize the discrepancy between the newly predicted outputs $C_{2,k}\hat{\textbf{x}}_{k|k-1}$ and the reference sensor outputs $\textbf{z}_{2,k}$ as an indication of the impact of unknown noises
\begin{align*}
\hat{\textbf{x}}_{k|k}&=\hat{\textbf{x}}_{k|k-1}+L_{k}(\textbf{z}_{2,k}-h_{2,k}(\hat{\textbf{x}}_{k|k-1}))
\end{align*}
where the state estimates are unbiased, i.e., ${\mathbb E}[\hat{\textbf{x}}_{k|k}]=\textbf{x}_k$, and the estimate gain matrix $L_k$ will be chosen such that the new estimates $\hat{\textbf{x}}_{k|k}$ have the smallest error variances.
Error dynamic and covariances are
\begin{align*}
\tilde{\textbf{x}}_{k|k}=\textbf{x}_k-\hat{\textbf{x}}_{k|k}= (I-L_{k}C_{2,k})\tilde{\textbf{x}}_{k|k-1}-L_k\xi_{2,k}
\end{align*}
and
\begin{align*}
P_{k}^x&=(I-L_kC_{2,k})P_{k|k-1}^x(I-L_kC_{2,k})^T + L_k R_{2,k} L_k^T\nonumber\\
&-(I-L_k C_{2,k}) G_{k-1}M_{2,k}R_{2,k}L_k^T\nonumber\\
&-L_kR_{2,k}M_{2,k}^TG_{k-1}^T(I-L_k C_{2,k})^T.
\end{align*}
To achieve optimal estimation, %the minimum covariance estimate, we use the trace norm of $P_{k}^x$ as an objective function of its 
we solve the variance minimization problem: $\min_{L_k} {\rm tr} (P_{k}^x)$.
%\begin{align*} \min_{L_k} {\rm tr} (P_{k}^x)%=\min_{L_k} {\rm tr} (P_{k}^{x*}+L_k\tilde{R}_k^*L_k^T-2L_kJ_k^T). \end{align*} This is a non-constrained convex optimization problem. Therefore, the gradient of the objective function will vanish at the minimizer.
We take the derivative of the objective function with respect to the decision variable $L_k$ and set it as zero
%\begin{align*} \frac{\partial {\rm tr} (P_{k}^x)}{\partial L_k}=2\tilde{R}_k^*L_k^T-2J_k^T=0. \end{align*} The solution is
\begin{align*}
L_k=(C_{2,k}P_{k|k-1}+R_{2,k}M_{2,k}^TG_{2,k-1}^T)^T\tilde{R}_{2,k}^{-1}
\end{align*}
where $\tilde{R}_{2,k} \triangleq C_{2,k}P_{k|k-1}^xC_{2,k}^T+R_{2,k} + C_{2,k}G_{k-1}M_{2,k}R_{2,k}+R_{2,k}M_{2,k}^TG_{k-1}^TC_{2,k}^T$ must be invertible.

\textbf{Testing sensor anomaly vector $\textbf{d}_{k}^{s}$ estimation}:
Given $\hat{\textbf{x}}_{k|k}$, the linear estimation for unknown sensor anomaly vector $\textbf{d}_{k}^s$ can be
\begin{align}
\hat{\textbf{d}}_{k}^s&= M_{1,k} (\textbf{z}_{1,k}- h_{1,k}(\hat{\textbf{x}}_{k|k}))\nonumber\\
&=M_{1,k}(C_{1,k} \tilde{\textbf{x}}_{k|k}+\textbf{d}_{k}^s +\xi_{1,k})
\label{P205}
\end{align}
where the estimates are unbiased, i.e., ${\mathbb E}[\hat{\textbf{d}}_{k}^s]=\textbf{d}_{k}^s$, providing that $M_{1,k}=I$. This also can be found by Gauss Markov theorem. By the theorem, the optimal estimates are 
\begin{align*}
M_{1,k} \triangleq ( \tilde{R}_{1,k}^{-1})^{-1}\tilde{R}_{1,k}^{-1}=I
\end{align*}
where $\tilde{R}_{1,k}=C_{1,k}P_{k}^x C_{1,k}^T + R_{1,k}$. The covariance matrices can be obtained by
\begin{align*}
P_{k}^{s}&=\tilde{R}_{1,k}
%\nonumber\\P_{k}^{xd_1}&=(P_{k}^{d_1x})^T=P_{k}^x C_{1,k}^TM_{1,k}^T.
%\label{P207}
\end{align*}

\textbf{Likelihood of a mode}:
In order to determine the ground truth condition of a robot, i.e., mode, we calculate a likelihood that reflects the discrepancy between the predicted output and the measured output of a mode.
For $\forall m$, we quantify the discrepancy between the predicted output and the measured output as follows
\begin{align*}
\nu_k^m = \textbf{z}_{2,k}-h_{2,k}^m(\hat{\textbf{x}}_{k|k-1}^m)).
\end{align*}
We approximate the output error $\nu_k^{m}$ as a multivariate Gaussian random variable. Then, the likelihood function is given by
\begin{align*}
{\mathcal N}_k^{m}
&\triangleq {\mathcal P}(y_k|m={\rm true})=
{\mathcal N}(\nu_k^{m};0,\bar{P}_{k|k-1}^{m})\nonumber\\
&=\frac{\exp(-(\nu_k^{m})^T(\bar{P}_{k|k-1}^{m})^{\dagger}\nu_k^{m}/2)}{(2 \pi)^{n^{m}/2} |\bar{P}_{k|k-1}^{m}|_{+}^{1/2}}
\end{align*}
where $\bar{P}_{k|k-1}^m = C_{2,k}^mP_{k|k-1}^{x,m}(C_{2,k}^m)^T + R_{2,k}^m-C_{2,k}^mG_{k-1}^mM_{2,k}^mR_{2,k}^m-R_{2,k}^m(M_{2,k}^m)^T(G_{k-1}^m)^T(C_{2,k}^m)^T$ is the error covariance matrix of $\nu_k^m$ and $n^m = Rank(\bar{P}_{k|k-1}^m)$.
Notations $\dagger$ and $|\cdot|_{+}$ refer to pseudoinverse and pseudodeterminant, respectively.
By the Bayes' theorem, the a posteriori probability is
$\mu_k^{m} \triangleq {\mathcal P}(m={\rm true}|y_k,\cdots,y_0) =\frac{{\mathcal P}(y_k|m={\rm true}){\mathcal P}(m={\rm true}|y_{k-1},\cdots,y_0)}{\sum_{i=1}^{\mathcal M}{\mathcal P}(y_k|m={\rm true}){\mathcal P}(m={\rm true}|y_{k-1},\cdots,y_0)}=\frac{{\mathcal N}_k^m\mu_{k-1}^m}{\sum_{i=1}^{\mathcal M}{\mathcal N}_k^m\mu_{k-1}^m}$.
However, such updates might cause the $\mu_k^m$ of certain modes to converge to zero.
To prevent this, we modify the posterior probability update to the following
\begin{align*}
\bar{\mu}_k^{m} = \frac{\mu_k^{m}}{\sum_{i=1}^{\mathcal M}\mu_k^i},
\end{align*}
where $\mu_k^{m} = \max\{{\mathcal N}_k^m\mu_{k-1}^{m},\epsilon\}$, and 
$\epsilon>0$ is a pre-selected small constant preventing the vanishment of the mode probability. 
The last step is to generate estimates of states and anomaly vector estimates of the maximum a posteriori mode.

\section{Khepera Dynamic Model}\label{sec:kin_mea_model}
\textbf{Kinematic model} The kinematic model of Khepera includes three states: $(x,y)$ is the robot location at a 2-D plane, and $\theta$ is its heading. The control commands are specified by two variables: $v_L$ and $v_R$, which are the speeds of the left and right wheels, respectively. %the velocity and angular velocity can be denoted as $v_k=\frac{v_L+v_R}{2}$ and $w_k=\frac{v_R-v_L}{r/2}$, where $r$ is the distance between the two wheels. For deduction simplicity purpose, we use the velocityhttps://www.sharelatex.com/project/5914d5ef91debba170c6fd46 and angular velocity as the inputs, and convert them back to $v_L$ and $v_R$ afterwards. 
Considering actuator misbehaviors with anomaly vector $\textbf{d}_{k-1}^a=[d_{k-1}^{a, L}, d_{k-1}^{a, R}]^T$ on the left and right wheel, the kinematic model can be presented as
\begin{align}
x_{k} &= x_{k-1} + T\cos \theta_{k-1}(v_L+d_{k-1}^{a, L}+v_R+d_{k-1}^{a, R})/2+\zeta_{k-1}^x \nonumber\\
y_{k} &= y_{k-1} + T\sin \theta_{k-1}(v_L+d_{k-1}^{a, L}+v_R+d_{k-1}^{a, R})/2+\zeta_{k-1}^y \nonumber\\
\theta_{k} &= \theta_{k-1} + T(v_R+d_{k-1}^{a, R}-v_L-d_{k-1}^{a, L})/\frac{D}{2}+\zeta_{k-1}^\theta
\label{ie9}
\end{align}
where $\zeta_{k-1}=[\zeta_{k-1}^x, \zeta_{k-1}^y, \zeta_{k-1}^\theta]^T$ is assumed to be zero mean Gaussian process noises, and $D$ is the distance between the left and right wheel on the chassis of Khepera. %Vectors $\textbf{x}_k = [x_k, y_k, \theta_k]^T$ and $\textbf{x}_{k+1} = [x_{k+1}, y_{k+1}, \theta_{k+1}]^T$ denote current states (position and orientation) and next states, respectively.% Velocity and angular velocity are denoted as $v_k=\frac{v_L+v_R}{2}$ and $w_k=\frac{v_R-v_L}{r/2}$. We specify $\textbf{u}_k = [v_k, \omega_k]^T$ as inputs, which is directly controllable from the planner.

\begin{figure}[!t]
  \centering
  \includegraphics[width =0.85\linewidth]{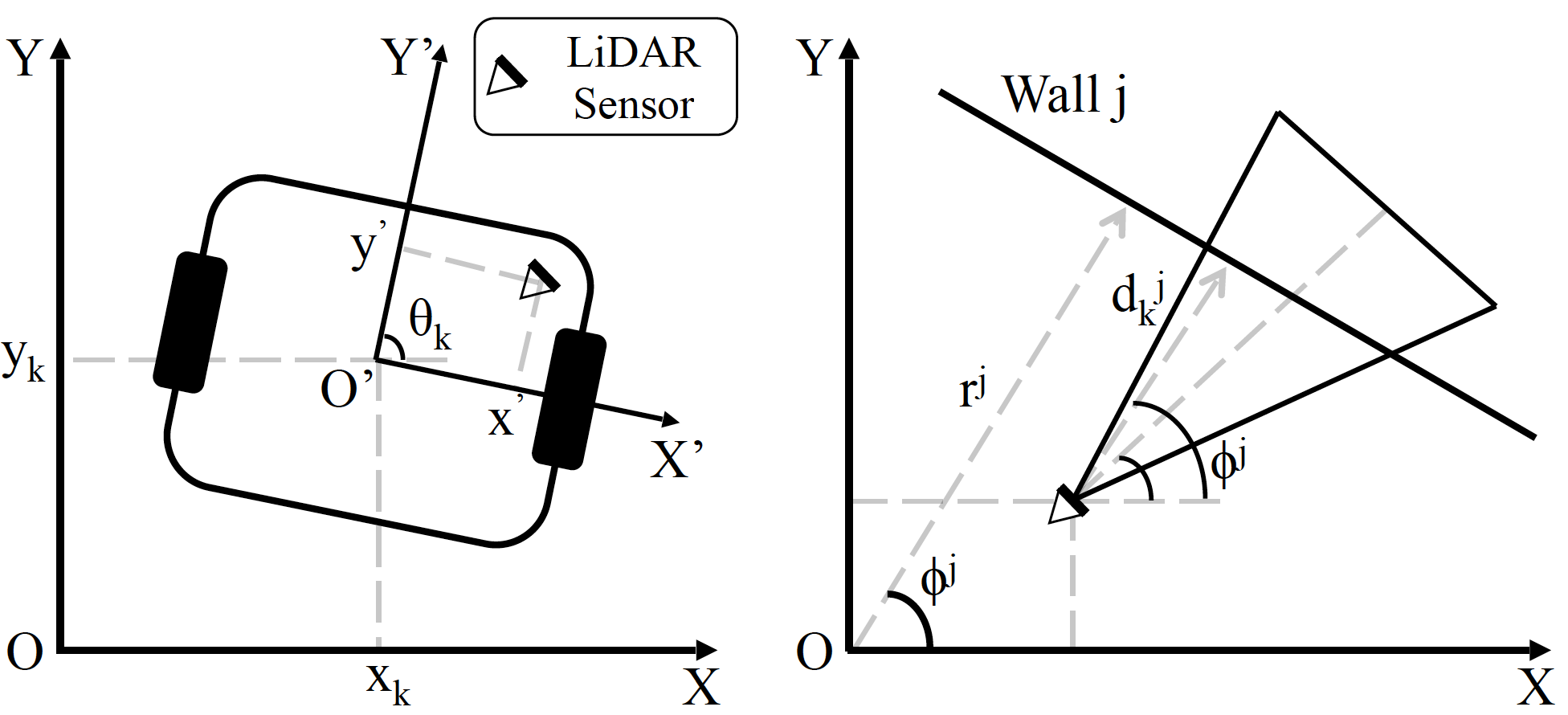}
  \caption{LiDAR sensor measurement model.}\label{khepera_3_math}
\end{figure}

\textbf{Measurement model} The sensor readings include sensing data from three sensors: $\textbf{z}_{k} = [\textbf{z}_{k, I}, \textbf{z}_{k, W}, \textbf{z}_{k, L}]^T$ where $\textbf{z}_{k, I}$ is from the IPS, $\textbf{z}_{k, W}$ is from the wheel encoder, and $\textbf{z}_{k, L}$ is from the LiDAR.

IPS sensor directly measures the states of Khepera, hence, the measurement model can be directly specified by
\begin{align}
\textbf{z}_{k, I} &= \textbf{x}_{k}+\textbf{d}_{k, I}^s+\xi_{k, I}
\label{eq:measurement_ips}
\end{align}
where $\xi_{k, I}=[\xi_{k, I}^x, \xi_{k, I}^y, \xi_{k, I}^\theta]^T$ refers to measurement noises from the IPS sensor, and $\textbf{d}_{k, I}^s=[d_{k,I}^{s, x},d_{k,I}^{s, y},d_{k,I}^{s, \theta}]$ refers to the sensor anomaly vector on IPS. %Suppose $\textbf{x}_{k-1} = [x_{k-1}, y_{k-1}, \theta_{k-1}]^T$ is the previous states of Khepera, the distance measurement from wheel encoder can be referred to as $\textbf{z}_k=[l_L+d_{k, W}^{sL}+\xi_{k, W}^{sL}, l_R+d_{k, W}^{sR}+\xi_{k, W}^{sR}]^T$, where $\xi_{k, W}=[\xi_{k, W}^L, \xi_{k, W}^R]^T$ refers to measurement noises from wheel encoder and $\textbf{d}_{k, W}^s=[d_{k, W}^{sL}, d_{k, W}^{sR}]^T$ refers to the sensor anomaly vector on the wheel encoder. The measurement model can be represented as:
% \begin{align}
% x_{k} &= x_{k-1} + (z_k(1)+z_k(2))\cos\theta_{k}/2 \nonumber\\
% y_{k} &= y_{k-1} + (z_k(1)+z_k(2))\sin \theta_{k}/2 \nonumber\\
% \theta_{k} &= \theta_{k-1} + (z_k(2)-z_k(1))/r
% \label{eq:measurement_we}
% \end{align}
\begin{figure}
  \begin{center}
  \includegraphics[width=0.43\columnwidth]{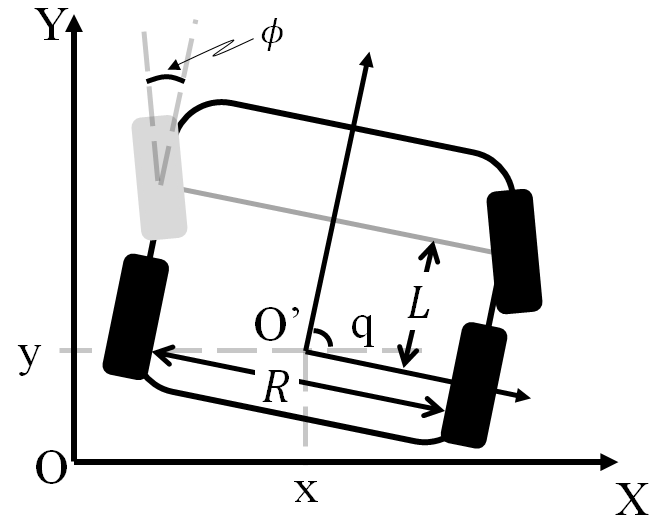}
  \end{center}
  \caption{Kinematic model of a rear-wheel-drive vehicle.}\label{fig:vehicle}
\end{figure}

The raw data measured by the wheel encoder are the distances traveled by each wheel $(l_L, l_R)$ in a control iteration. For convenience reasons, we convert them into robot states using previous states $\textbf{x}_{k-1}$ before we feed the data to the planner
\begin{align}
x_{k} &= x_{k-1} + (l_L+l_R)\cos\theta_{k}/2 \nonumber\\
y_{k} &= y_{k-1} + (l_L+l_R)\sin \theta_{k}/2 \nonumber\\
\theta_{k} &= \theta_{k-1} + (l_R-l_L)/r \nonumber
\end{align}
Analogously with IPS, the measurement model for the wheel encoder can be specified as
\begin{align}
\textbf{z}_{k, W} &= \textbf{x}_{k}+\textbf{d}_{k, W}^s+\xi_{k, W}
\label{eq:measurement_ips}
\end{align}
after the conversion, where $\xi_{k, W}=[\xi_{k, W}^x, \xi_{k, W}^y, \xi_{k, W}^\theta]^T$ refers to measurement noises from the wheel encoder and $\textbf{d}_{k, W}^s=[d_{k,W}^{s, x},d_{k,W}^{s, y},d_{k,W}^{s, \theta}]^T$ refers to the sensor anomaly vector on the wheel encoder.

The LiDAR sensor is placed on top of the robot with a shift distance of $[x',y']^T$ from the origin $O'$ as shown in the left plot of Figure~\ref{khepera_3_math}. Raw sensor readings returned from LiDAR are the distances between LiDAR and the surrounding walls (see the right plot of Figure~\ref{khepera_3_math}). Given the LiDAR readings, we process the raw data into the perpendicular distance $l^{j}_k$ from each boundary wall $j \in \{1, 2, 3, 4\}$ and the orientation $\theta_{k}$of Khepera. Specifically, we recognize the straight line segments using raw distances from all direction, and calculate the distances to each wall as follows
\begin{align}
  l^{j}_k %&= r^j - x_k \cos \phi^j - y_k \sin \phi^j+ \xi_k \nonumber\\
  & =  r^j - (x_k+ x'\sin \theta_k + y' \cos \theta_k) \cos \phi^j\nonumber\\
  & \ \ - (y_k - x'\cos \theta_k + y' \sin \theta_k) \sin \phi^j +d_{k,L}^{s,j} +\xi_{k, L}^j
  \label{ee001}
\end{align}
where $\xi_{k, L}=[\xi_{k, I}^j]^T,j \in \{1, 2, 3, 4\}$ refers to measurement noises from LiDAR. The distance $r^j$ and the angle $\phi^j$ of each wall in the global coordinate is known in advance as the map information. Using $\phi^j$ of each wall and the 240 degrees of range, we can also infer the angle of the robot. We use the distance and the angle to each wall as the sensor readings from LiDAR: $\textbf{z}_{k, L} =[l_{k,}^{j}, \theta_k]^T,j \in \{1, 2, 3, 4\} $. In outdoor environments, LiDAR measurement model can be obtained using more complicated simultaneous localization and mapping (SLAM) algorithms~\cite{durrant2006simultaneous}. For demonstration purposes, we apply a simple transformation in the indoor environment~\cite{jetto1999development}.

\section{Tamiya RC Car Dynamic Model}
\textbf{Kinematic model} The kinematic model of a Tamiya RC car is presented in Figure~\ref{fig:vehicle}. The states of the vehicle also include the location and the orientation $(x,y,\theta)$ in a 2D plane. The control includes the longitudinal velocity and the steering $(v,\phi)$ . The kinematic model of the vehicle can be described as
\begin{align*}
x_{k} &= x_{k-1} + T(v_{k-1}+d_{k-1}^v) \cos \theta_{k-1} + \zeta_{k-1}^x\nonumber\\
y_{k} &= y_{k-1} + T(v_{k-1}+d_{k-1}^v)\sin\theta_{k-1}+\zeta_{k-1}^y\nonumber\\
\theta_{k} &= \theta_{k-1}+T \frac{v_{k-1}}{L}\tan (\phi_{k-1}+d_{k-1}^{\phi})+\zeta_{k-1}^{\theta}
\end{align*}
where $\zeta_{k-1}=[\zeta_{k-1}^x, \zeta_{k-1}^y, \zeta_{k-1}^\theta]^T$ is assumed to be a zero mean Gaussian process noise vector, $\textbf{d}_{k-1}^a=[d_{k-1}^v,d_{k-1}^{\phi}]^T$ is the actuator anomaly vector, $L$ is the wheelbase, and $T$ is the control iteration interval.

\textbf{Measurement model} At each instant of time, sensor readings include data from three sensors: $\textbf{z}_{k} = [\textbf{z}_{k, I}, \textbf{z}_{k, W}, \textbf{z}_{k, M}]^T$, where each vector refers to the sensor readings from IPS, LiDAR, and IMU, respectively. The measurement models for IPS and LiDAR are similar to those in Khepera (see Section~\ref{sec:kin_mea_model}).

The IMU sensor generates a quaternion $[q_0,q_1,q_2,q_3]^T$, a 3-D acceleration $\textbf{a}_{k,M}^{local}$,
and a 3-D rotational speed $\textbf{w}_{k,M}^{local}$ on a body-fixed coordinate.
We first obtain the coordinate transformation matrix $C(q)$ from the body-fixed coordinate to the global coordinate~\cite{kuipers1999quaternions}.
% \begin{strip}
\begin{small}
\begin{align*}
&C(q)=\\
&\left[
\begin{array}{ccc}
q_0^2+q_1^2-q_2^2-q_3^2&2(q_1q_2-q_0q_3)&2(q_1q_3+q_0q_2)\\
2(q_1q_2+q_0q_3)&q_0^2-q_1^2+q_2^2-q_3^2&2(q_2q_3-q_0q_1)\\
2(q_1q_3-q_0q_2)&2(q_2q_3+q_0q_1)&q_0^2-q_1^2-q_2^2+q_3^2\\
\end{array}
\right].
\end{align*}
\end{small}
% \label{eq:coord}
% \end{strip}
The acceleration vector and the rotation speed on the global coordinate system can be obtained as $C(q)\textbf{a}_{k,M}^{local}$ and $C(q)\textbf{w}_{k,M}^{local}$, respectively.
% \begin{align*}
% \textbf{a}_{k,M}^{global}&=[a_{k,M}^x,a_{k,M}^y,a_{k,M}^z]^T = C(q)\textbf{a}_{k,M}^{local}\nonumber\\
% \textbf{w}_{k,M}^{global}&=[w_{k,M}^x,w_{k,M}^y,w_{k,M}^z]^T = C(q)\textbf{w}_{k,M}^{local}.
% \end{align*}
The vehicle velocity vector can be updated by: $\textbf{v}_k=[v_{k,M}^x,v_{k,M}^y,v_{k,M}^z]^T = \textbf{v}_{k-1}+\textbf{a}_k^{global}T$.
Then the state vector can be calculated by integration as follows%: $x_k=x_{k-1}+v_{k,M}^xT+\frac{1}{2}a_{k,M}^x T^2$, $y_k=y_{k-1}+v_{k,M}^yT+\frac{1}{2}a_{k,M}^y T^2$, $\theta_k = \theta_{k-1}+w_{k,M}^z T$.
\begin{align*}
x_k&=x_{k-1}+v_{k,M}^xT+\frac{1}{2}a_{k,M}^x T^2\nonumber\\
y_k&=y_{k-1}+v_{k,M}^yT+\frac{1}{2}a_{k,M}^y T^2\nonumber\\
\theta_k &= \theta_{k-1}+w_{k,M}^z T.
\end{align*}

\section{Separating Actuator Anomaly Vector}\label{sec:separation}
%correlated covariance problem
In Section IV.D. of~\cite{dsn_paper}, we mention that \texttt{RoboADS} only checks the aggregate test statistics instead of each individual actuator. This section explains the reason in detail.

At a high level, the actuator anomaly vectors are statistically correlated. Without loss of generosity, we consider a robot with two actuators such as Khepera. During actuator anomaly vector estimation, we obtain $\hat{\textbf{d}}_k^a =[\hat{d}_k^L, \hat{d}_k^R]^T$, with error covariances $P_k^a$. In Algorithm 1 line 20, we test 
\begin{align}
(\hat{\textbf{d}}_k^a)^T (P_k^a)^{-1} \hat{\textbf{d}}_k^a \geq \chi_{p=2}(\alpha)
\label{ttp0}
\end{align}
to determine the existence of actuator misbehaviors. The threshold $\chi_{p=2}(\alpha)$ is a Chi-square test value with the degree of freedom $p=2$ and the confidence level $\alpha$.

In order to confirm actuator misbehaviors on each actuator, we need to separately conduct Chi-square test $\hat{d}_k^L$, and $\hat{d}_k^R$, with corresponding marginal variances $P_k^a(1,1)$, and $P_k^a(2,2)$:
\begin{align}
(\hat{d}_k^L)^T (P_k^a(1,1))^{-1} \hat{d}_k^L \geq \chi_{p=1}^2(\alpha)\nonumber\\
(\hat{d}_k^R)^T (P_k^a(2,2))^{-1} \hat{d}_k^R \geq \chi_{p=1}^2(\alpha).
\label{ttp1}
\end{align}
However, a positive testing result in~\eqref{ttp0} does not guarantee a positive testing result in~\eqref{ttp1} because the off-diagonal terms of matrix $P_k^a$ are neglected in~\eqref{ttp1}. The explanation is shown as follows:
\begin{align}
&(\hat{\textbf{d}}_k^a)^T (P_k^a)^{-1} \hat{\textbf{d}}_k^a = (\hat{d}_k^L)^T (P_k^a)^{-1}(1,1) \hat{d}_k^L\nonumber\\
&\quad \quad \quad +(\hat{d}_k^L)^T (P_k^a)^{-1}(1,2) \hat{d}_k^R
+(\hat{d}_k^R)^T (P_k^a)^{-1}(2,1) \hat{d}_k^L\nonumber\\
&\quad \quad \quad +(\hat{d}_k^R)^T (P_k^a)^{-1}(2,2) \hat{d}_k^R\nonumber\\
&(\hat{d}_k^L)^T (P_k^a(1,1))^{-1} \hat{d}_k^L = (\hat{d}_k^L)^T (P_k^a(1,1))^{-1} \hat{d}_k^L\nonumber\\
&(\hat{d}_k^R)^T (P_k^a(2,2))^{-1} \hat{d}_k^R = (\hat{d}_k^R)^T (P_k^a(1,1))^{-1} \hat{d}_k^R
\end{align}
%We cannot compare~\eqref{ttp0} and~\eqref{ttp1} directly since $(P_k^a(1,1))^{-1} \neq (P_k^a)^{-1}(1,1)$.
Note that $(\hat{\textbf{d}}_k^a)^T (P_k^a)^{-1} \hat{\textbf{d}}_k^a=(\hat{d}_k^L)^T (P_k^a(1,1))^{-1} \hat{d}_k^L+(\hat{d}_k^R)^T (P_k^a(2,2))^{-1} \hat{d}_k^R$ if $P_k^a$ is a diagonal matrix.

% nonlinearity of Chi-square threshold
%Even though $P_k^a$ is a diagonal matrix so that we could separate them, there is another problem that Chi-square test threshold is nonlinear; i.e., 
Another problem for the separation is that the Chi-square test threshold is nonlinear. For instance, 
$\chi_{p=1}^2(0.01)=6.635$ and $\chi_{p=2}^2(0.01)=9.210$.
Suppose $P_k^a$ is a diagonal matrix and the test scores after separation are $(\hat{d}_k^L)^T (P_k^a(1,1))^{-1} \hat{d}_k^L=5$ and $(\hat{d}_k^R)^T (P_k^a(2,2))^{-1} \hat{d}_k^R=5$. The actuator misbehaviors would be detected by~\eqref{ttp0} but not by~\eqref{ttp1}.

Therefore, we conduct the Chi-square test on the aggregate actuator anomaly vector instead of the separated vector components. The decision results from the hypothesis tests indicate whether the robot has actuator misbehaviors with a certain level of confidence, yet no decision is made on whether a particular actuator is misbehaving.
\bibliographystyle{IEEEtran}
\bibliography{biblio}

\end{document}